\documentclass[preprint,showpacs,amsmath,amssymb,pre]{revtex4}

\usepackage{graphicx}
\usepackage{dcolumn}
\usepackage{bm}

\newcommand{\partN}[3]{\frac{\partial^{#3} {#1}}{\partial {#2}^{#3}}}

\newcommand{\partopN}[2]{\frac{\partial^{#2}}{\partial {#1}^{#2}}}
\newcommand{\partop}[1]{\frac{\partial}{\partial {#1}}}

\newtheorem{theorem}{Theorem}

\begin{document}


\title{Trapping and Steering on Lattice Strings:\\ Virtual Slow
Waves, Directional and Non-propagating Excitations}

\author{Georg Essl}
\email{georg@mle.media.mit.edu}
 \affiliation{%
Media Lab Europe\\
Sugar House Lane\\
Dublin 8, Ireland}

\date{\today}

\begin{abstract}
Using a lattice string model, a number of peculiar excitation
situations related to non-propagating excitations and non-radiating
sources are demonstrated. External fields can be used to trap
excitations locally but also lead to the ability to steer such
excitations dynamically as long as the steering is slower than the
field's wave propagation. I present explicit constructions of a number
of examples, including temporally limited non-propagating excitations,
directional excitation and virtually slowed propagation. Using these
dynamical lattice constructions I demonstrate that neither persistent
temporal oscillation nor static localization are necessary for
non-propagating excitations to occur.
\end{abstract}

\maketitle

\section{Introduction}

Can a local excitation (source) in classical field theories be
invisible to observers outside the region of excitation? This
question has recently received renewed interest. 

Berry {\em et al.} \cite{BFGW98} described a peculiar excitation case
for the one-dimensional wave-equation of a perfectly elastic string
under tension. They show that the response of the string can be made
to be confined to a bounded region by carefully choosing a forced
excitation of oscillatory type. This means that the excitation will
not propagate away along the string. Denardo gives a simple and
intuitive explanation by using a wave interference argument
\cite{Denardo98}. Gbur {\em et al} \cite{GFW99} discuss conditions of
finite string length and dissipation.

Other recent work investigated non-propagating excitations include
Marengo and Ziolkowski \cite{MZ99,MZ00a,MZ00c} who discuss the
generalization of non-propagating conditions of D'Alembertian
($\Box\stackrel{\text{\tiny def}}{=}\nabla^2-c^{-2}\partopN{t}{2}$)
operators and its temporally reduced version the Helmholtz operator
($\nabla^2+k^2$) on various related classical scalar and vector
fields. Marengo, Devaney and Ziolkowski \cite{MDZ99} give the
condition for time-dependent but not necessarily time-harmonic
non-radiating sources and for selective directional radiation for the
inhomogeneous wave equation in three spatial dimensions. Marengo and
Ziolkowski \cite{MZ00b} generalize these conditions to more general
scalar and vector field dynamics. Marengo, Devaney and Ziolkowski
\cite{MDZ00} also give examples in one and three spatial dimension
for the time-harmonic case. Hoenders and Ferwerda \cite{HF01} discuss
the relationship of non-radiating and radiating parts of the case of
the reduced Helmholtz equation, which can be derived from the string
equation by assuming general oscillatory time solutions (see
\cite{BFGW98}). Denardo and Miller \cite{DM03} discuss the related
case of leakage from an imperfect non-propagating excitation on a
string. Gbur \cite{Gbur03} provides a comprehensive recent review of
this topic and the reader is referred to this review for more detailed
historical context. Of the earlier work the following contributions
are particularly relevent for the discussion here: Schott
\cite{Schott33,Schott37} gave the condition for non-radiation of a
spherical shell on a circular orbit. Bohm and Weinberg \cite{BW48}
extended this result to more general spherical charge distributions
and Goedecke \cite{Goedecke64} showed how an asymmetrical charge
distribution with spin is non-radiating. All of these works are
concerned with the case of spatially moving sources. Finally it is
noteworthy, that non-radiating sources play an important role in
inverse problems and have been investigated in a one-dimensional
electrodynamic situation by Habashy, Chow and Dudley \cite{HCD90}.

In this paper our purpose is to describe this phenomena in the 
case of a lattice string in one dimensions by discretizing
D'Alembert's solution. This approach is used extensively to
simulate vibrating strings and air tubes of musical instruments. See
\cite{Smith97} and references therein.


This leads to explicit dynamical constructions of previously reported
non-propagating excitations. Its simplicity allows for additional
insight into the mechanism that allows for the local confinements and
the conditions under which they occur. I will show how the basic
mechanisms that provide a time-harmonic stationary non-propagating
excitation in one dimension as studied by Berry {\em et al.} and Gbur
{\em et al.}  \cite{BFGW98,GFW99} allows for a much wider class of
excitations. For instance can such an excitation be relieved from the
time-harmonic assumption beyond one period allowing for
non-propagating sources that are short-lived. Directional excitations
can easily be achieved using very simple bidirectional excitation
patterns. These are explicit constructions of such waves in one
spatial dimension whose general condition of existence in the
three-dimensional case has been derived by Marengo, Devaney and
Ziolkowski \cite{MDZ99}. Wave propagation can be virtually slowed
down. In general I will show that non-propagating excitations can be
extended to steered excitation regions with basic physical
restrictions imposed by the underlying field dynamics.

First I will give a quick derivation of the simple lattice model from
the wave equation as can also be found in \cite{Smith97}. Then I will
give a new argument and construction of the Berry {\em et al.} type
non-propagating excitation purely based on discrete string
dynamics. This will then be compared to the original approach. Then I
will extend the discussion to examples of additional types of
non-propagating waves, including directional and slowed waves. Finally
I will discuss very general constraints on such ``steered'' localized
excitations.

\section{Lattice String Model}

The lattice string model can easily be derived from the
wave-equation by discretizing the D'Alembert solution. Hence the
continuous case will be discussed first.

\subsection{Continuous Wave Solutions}

The one-dimensional homogeneous wave equation of the perfectly elastic string
under tension is:

\begin{equation}\label{eq:eb}
\mu \partN{y}{t}{2}-K\partN{y}{x}{2}=0
\end{equation}

where $c^2=K/\mu$ is derived from mass density $\mu$ and tension
$K$. The D'Alembert solution of the homogeneous ``free field'' case
has the well known form \cite[p. 596, eq. (4)]{Kreyszig99}:

\begin{equation}\label{eq:we}
y(x,t)=w^+(x-ct)+w^-(x+ct)
\end{equation}

Hence the solution of the general of the homogeneous wave-equation are
two propagating waves whose content is restricted by initial and
boundary conditions.
As wave-equation is linear we have a connection
between initial conditions and external driving forces.  Driving
forces can be seen as infinitesimal time frames that act on the wave
dynamics by imposing a new initial condition at each point in
time. Hence we need to consider the initial value problem to gain
insight into both processes at once.

At a given time frame $t_i$ let the following initial conditions hold:

\begin{align}
y(x,t_i)&=f(x,t_i)\label{eq:in1}\\
y_t(x,t_i)&=g(x,t_i)\label{eq:in2}
\end{align}

Equation (\ref{eq:in1}) with (\ref{eq:we}) gives a particular solution $v^+$:

\begin{align}\label{eq:satin1}
v^+(x-ct_i)+v^-(x+ct_i)&=f(x,t_i)
\end{align}

Taking the first temporal derivative of (\ref{eq:we}) and satisfying equation (\ref{eq:in2}) we get:

\begin{align}
-cv_t^+(x-ct_i)+cv_t^-(x+ct_i)&=g(x,t_i)
\end{align}

Integrating with respect to $x$ we get \cite[eq. (10) p. 596]{Kreyszig99}:

\begin{align}\label{eq:satin2}
\begin{split}
-cv^+(x-ct_i)+cv^-(x+ct_i)=k(x_0)+\int_{x_0}^{x}g(s)ds,\\k(x_0)=-cv^+(x_0)+cv^-(x_0)
\end{split}
\end{align}

From equations (\ref{eq:satin1}) and (\ref{eq:satin2}) we can solve
for the traveling wave components:

\begin{align}
\begin{split}\label{eq:force1}
v^+(x-ct_i)=\frac{1}{2}f(x,t_i)-\frac{1}{2c}\int_{x_0}^{x-ct_i}&g(s)ds\\&-\frac{1}{2}k(x_0)
\end{split}\\\
\begin{split}\label{eq:force2}
v^-(x+ct_i)=\frac{1}{2}f(x,t_i)+\frac{1}{2c}\int_{x_0}^{x+ct_i}&g(s)ds\\&+\frac{1}{2}k(x_0)
\end{split}
\end{align}

We see that forced displacement $f(\cdot)$ splits evenly between left
and right traveling waves and the integrated forced velocity
$g(\cdot)$ splits with a sign inversion.

For our current discussion I will share the assumption of no initial
velocity of Berry {\em et al.} \cite{BFGW98} and hence the integral
over $g(\cdot)$ will vanish.

For the infinite string this is already the complete solution for any
twice differentiable function of free solutions and external forced
displacements.

\subsection{Discrete Wave Solutions}

To arrive at lattice equations we discretize the solution of the
wave-equation (\ref{eq:we}) in time via the substitution $t
\rightarrow Tn$ where $T$ is the discrete time-step and $n$ is the
discrete time index. This automatically corresponds to a
discretization in space as well, because in finite time $T$ a wave
will travel $X=cT$ distance according to (\ref{eq:we}). The spatial
index will be called $m$. The free-field discrete D'Alembert solution:

\begin{equation}\label{eq:wga}
\begin{split}
y(mX,nT)=&w^+(mX-cnT)+w^-(mX+cnT)
\end{split}
\end{equation}

In general we can always express all discrete equations in terms of
finite time steps or finite spatial lengths. We chose a temporal
expression and substitute $X=cT$ and suppress shared terms in $cT$ to arrive
at the index version of the discrete D'Alembert solution \cite{Smith97}:

\begin{equation}\label{eq:wg}
y(m,n)=w^+(m-n)+w^-(m+n)
\end{equation}

By equations (\ref{eq:force1}) and (\ref{eq:force2}) we see that at an
instance $m_i,n_i$ the discrete contribution of external forced
displacements splits evenly between the traveling waves and we arrive
at the discrete field equations including external forced
displacements:

\begin{align}
W^+(m_i-n_i)=w^+(m_i-n_i)+\frac{1}{2}f(m_i,n_i)\label{eq:fullwg1}\\
W^-(m_i-n_i)=w^-(m_i-n_i)+\frac{1}{2}f(m_i,n_i)\label{eq:fullwg2}
\end{align}

\section{Non-propagating Excitation}

Next we will construct the non-propagating excitation from the
lattice string dynamics directly. 

For simplicity and without loss of generality, we will assume a region
aligning with the discretization domain throughout. We want to
construct an excitation which is confined to a length $-L\leq x\leq
L$. For now we will assume that the string should otherwise stay at
rest. This implies that there are no incoming waves into the region
$\Omega=[-L,L]$ from the outside. We are interested in a non-trivial
excitation within the region.

First we consider the contributions to the position $-L$. As there are
no incoming external waves we get:

\begin{align}\label{eq:noin}
w^+(-L+n)=0
\end{align}

We do expect non-trivial wave $w^-(-L-n)$ to reach the boundary but we
require the total outgoing wave to vanish we have:

\begin{align}\label{eq:noout}
W^-(-L-n)=w^-(-L-n)+\frac{1}{2}f(-L,n)=0
\end{align}

The necessary external forced displacement contribution to for
cancellation needs to be:

\begin{align}\label{eq:forceeq}
\frac{1}{2}f(-L,n)=-w^-(-L-n)
\end{align}

The complete incoming wave (\ref{eq:fullwg1}) will see the same forced
contribution (\ref{eq:forceeq}) and with equation (\ref{eq:noin}) we
get:

\begin{align}\label{eq:cond1}
W^+(-L+n)=\frac{1}{2}f(-L,n)=-w^-(-L-n)
\end{align}

Hence the matched forced displacement leads to a reflection with sign
inversion at the region boundary at $-L$.

Following the same line of argument at point $L$ we get the related
condition:

\begin{align}\label{eq:cond2}
W^-(L-n)=\frac{1}{2}f(L,n)=-w^+(L+n)
\end{align}

With these two conditions we can study the permissible form of
excitations. First we assume an initial forced displacement impulse
from a position $p$ in the interior of the domain
$\Omega\setminus\partial\Omega=(-L,L)$. Hence $-L<p<L$ and
$f(p,0)=a_p$ with $a_p\in\mathbb{R}$.

It will take half the impulse $L+p$ steps to reach the left boundary
and the other half $L-p$ steps to reach the right one.

At each boundary the respective condition
(\ref{eq:cond1},\ref{eq:cond2}) needs to be satisfied and we get:

\begin{align}
f(-L,L+p)&=-f(p,0)\\
f(L,L-p)&=-f(p,0)
\end{align}

The impulse will then reflect back and create periodic matching conditions.

\begin{align}
f(-L,L+p+4L\nu)&=f(p,0)\label{eq:boundf1}\\
f(-L,L-p+(2\nu-1)2L)&=-f(p,0)\\
f(L,L-p+4L\nu)&=f(p,0)\\
f(L,L+p+(2\nu-1)2L)&=-f(p,0)\label{eq:boundf4}
\end{align}

with $\nu=1,2,\cdots$.

Hence we see that a single impulse will necessitate an infinite
periodic series of forced external displacements at the boundaries to
trap the impulse inside as each ``annihilation'' of a half-pulse
reaching the boundary leads to a ``creation'' of a reflected one.

The required impulse response of a boundary forced function $f(\pm
L,\cdot)$ can easily be observed from equations
(\ref{eq:boundf1}--\ref{eq:boundf4}) to be spatially periodic in $4L$ with an
initial phase factor dictated by the starting position
$p$. Additionally the functional shape of the impulse responses $f(\pm
L,\cdot)$ is completely defined for all time steps as $f(\pm L,
\cdot)=0$ for all times that equations (\ref{eq:boundf1}--\ref{eq:boundf4}) don't
apply.

A condition for stopping a non-propagating excitation can be derived
from the fact that a impulse will return to its initial position every
$4L$ time steps. Additionally it is easy to see that the traveling
impulses will occupy the same spatial position every odd multiple of
$2L$ with a sign inversion. Hence an impulsive forced displacement
$f((-1)^{\mu-1}p,4L\mu)=(-1)^{\mu-1}a_p$ with $\mu=1,2,\cdots$ will cancel
an initial impulse $f(p,0)=a_p$. From this we can immediately deduce
the following property:

\begin{theorem}
The shortest possible single impulse finite non-propagating excitation
takes $2L$ time-steps.\label{th:shortest}
\end{theorem}

and more generally:

\begin{theorem}
The time of any single impulse excitation finite non-propagating
excitation has to be $2\mu L, \mu\in\mathbb{N}$.
\end{theorem}

More importantly we observe the property: {\em Non-propagating excitations can be finite in duration}.

This is an extension beyond Berry {\em et al.} \cite{BFGW98} which assumes
infinitely periodic temporal progressions in their derivations.

The general solution for discrete non-propagating wave functions can
be derived by observing that any initial ``phase'' $p_i$ is orthogonal
to other phases $p_j$ for $i,j \in
\Omega\setminus\partial\Omega=(-L,L)$, i.e. $\langle
f(p_i,0),f(p_j,0)\rangle =0$ for $i\neq j$. Within a $2L$ period
$f(\pm L,\cdot)$ is well-defined by $\sum_i
f(p_i,\cdot)$. Interestingly though this provides the only restriction
to the forced boundary functions. This can be seen by Theorem
\ref{th:shortest}. After $2L$ each $p_i$ will find constructive
interference and can be annihilated or rescaled to an arbitrary other
value $a_i(2L)$. Hence any arbitrary succession of $2L-2$ force
distributions with a $2L$ termination is permissible. Hence periodicity
is not necessary.

The time harmonic case can be derived if the initial force
distribution within the domain is not modified over time. Then a
configuration will repeat after traveling left and right, being
reflected at the domain boundary twice, traversing the length of the
region twice. Hence the lowest permissible wave-length is $4L$. By
reflecting twice the wave will have gone through a $2\pi$ phase shift,
but we note that the periodicity condition is also satisfied if any
number of additional $2\pi$ shifts have been accumulated. Hence we get
for permissible wave-numbers:

\begin{align}
k=\frac{2\pi n}{4L},\qquad\text{where}\ n=1,2,\cdots.
\end{align}

or

\begin{align}\label{eq:wavenumber}
kL=\frac{n\pi}{2}
\end{align}

By allowing only even $n$ we get the Berry {\em at al.}  condition
\cite{BFGW98} for an even square distribution. The odd $n$ situation
corresponds to the odd-harmonic out-of-phase construction proposed by
Denardo \cite{Denardo98}.

Many of these properties can be seen visually in the numerical
simulation depicted in Figure \ref{fig:nonprop}. 

It is interesting to observe that two synchronous point-sources
oscillating with the above phase condition will not be completely
non-propagating. They will only be non-propagating after waves created
at the wave onset have escaped. This is a refinement of the argument
put forward by Denardo \cite{Denardo98} and can intuitively be
described as {\em non-interference of the first trap period}. Hence
the first pairs of pulses will have half-Amplitude components escaping
in either direction but every subsequent period will be trapped. This
behavior, which could be called imperfect trapping or trapping with
transient radiation, is depicted in Figure
\ref{fig:pointpair}. Sources presented by Berry {\em et al.} and
Denardo \cite{BFGW98,Denardo98} do not display this behavior because
the force is assumed to be oscillatory at all times and hence has no
onset moment.

Non-propagating excitations
can be used as generic building blocks for other unusual excitation
induced behavior on the string. In particular I will next describe how
to construct an uni-directional emitter, and a virtually slowed
propagation. In fact a non-propagating excitation can be seen as
virtually stopping a wave at a particular position.

\section{Directional Excitations}

A one-sided open trap immediately suggests another unusual excitation
type, namely the directional excitation. The string is to be excited in
such a way that a traveling wave in only one direction results.

We start with a one-sided open trap. This is a trap that uses a
reflection condition (\ref{eq:cond1}) and (\ref{eq:cond2}) only on one
side of an initial excitation. Evidently the wave then can only
travel in the opposite direction. For the discussion we will describe
a right-sided propagator (i.e. a propagator traveling with increasing
negative index). The trapping condition then reads:

\begin{align}
f(m+1+p,n+p)=-f(m,n-1)
\end{align}

Hence the trapping excitation point is a $p$ time-step lagging negative
copy of the original excitation. The emitted wave will have the form

\begin{align}
\frac{1}{2}f(m+1,n+2p)-\frac{1}{2}f(m+1,n)
\end{align}

The emitting wave will show self-interference at a phase of $2p$
time-steps, as can be seen in the simulation depicted in Figure
\ref{fig:onesided}. In general the self-interference phase can be
chosen by the distance $p$ between the wave creation point and the
trapping point. It is worth noting that it is possible to eliminate
interference by trapping the lagging contribution and hence create a
wave non-interference directional wave left of the trapping region.

\section{Virtual Slow Waves}

Virtual slow waves can be achieved by alternating directional wave
propagation with trapping. The slowness of the wave propagation can be
controlled by the number and and duration times of the traps along a
propagation. The propagation characteristics of the dynamic operator
has not changed at all, hence we call the this state ``virtually
slow'' as opposed to the case where the field itself induces a change
in wave propagation speed. This also means that within a slowed or
``steered'' region the wave propagation is the one prescribed by the
dynamic operator
$(\partop{x}+c\partop{t})(\partop{x}-c\partop{t})$ on the string $y(x,t)$.

The amount of time spend in traps determines the overall slowness. One
example of slow wave consists of an immediate alteration between one
stage of trapping and one step of one-sided propagation is illustrated
in Figure \ref{fig:slowwave}. The effective propagation speed of the
wave can easily be read from the diagram to give
$c_{\text{eff}}=c\frac{X}{3T}=c/3$. As is evident from Theorem
\ref{th:shortest}, a unit $L=1$ trap will last $2$ time-steps and will
not propagate spatially and one step of free propagation will last one
time-step and and make one spatial step, hence resulting in a spatial
to temporal ratio of $1:3$.

The trapping relations are:

\begin{align}
\begin{split}
f(m-2-\nu,n+1+6\nu)&=f(m+1-\nu,n+6\nu)\\
&=-f(m,n-1)
\end{split}\\
\begin{split}
f(m-3-\nu,n+4+6\nu)&=f(m-\nu,n+3+6\nu)\\&=f(m,n-1)
\end{split}
\end{align}

with $\nu=0,2,4,\cdots$.

\section{Steering}

The generalized interpretation of the excitation interaction lead to
the general dynamical confinement of waves by external excitation. For
instance following very similar arguments as for virtual slow waves a
construction is possible which gives a slowed ``cone of influence'' by
successively widening the trap boundaries at a speed slower than the
the wave speed $c$. By this argument it is sufficient for the trap
boundaries' change to be less than $c$ for it to be trapping the
wave. This is not a necessary condition by the following
counter-example: Let the trap width be $L$ and change rapidly by some
slope $dL>c$ to some new constant width $L_2$ at which it becomes
constant. Obviously the wave will then be able to reach the new
boundary even though a local change of the boundary exceeded the
dynamical speed $c$. The necessary condition can be seen from our
previous construction. At a trap boundary a wave is reflected and will
propagate in the opposite direction of the domain following the linear
characteristic $c$. Only if this characteristics intersects with the
dynamic trapping boundary will there be another externally forced
reflection as illustrated in Figure \ref{fig:steering}. These may
in fact have regions where no trapping is necessary and possible.

\section{Interaction with Background Fields}

It is important to note that while we assumed that the incoming wave
vanishes, see equation (\ref{eq:noin}), the outgoing wave condition
(\ref{eq:noout}) does not change if there is in fact an incoming
wave. The ``reflection wave'' (\ref{eq:cond1}) and
(\ref{eq:cond2}) can be rewritten for a non-zero incoming field
without affecting the trapping:

\begin{align}
\begin{split}
W^+(-L+n)&=w^+(-L+n)+\frac{1}{2}f(-L,n)\\
\frac{1}{2}f(-L,n)&=-w^-(-L-n)
\end{split}
\end{align}
and
\begin{align}
\begin{split}
W^-(L-n)&=w^-(L-n)+\frac{1}{2}f(L,n)\\
\frac{1}{2}f(L,n)&=-w^+(L-n)
\end{split}
\end{align}

These conditions are ``absorbing'' in the sense that an external
field entering the trapping region will not leave it.

The ``non-interacting'' property of a trap defined by the periodic
matching conditions (\ref{eq:boundf1}--\ref{eq:boundf4}) can be seen by
assuming a non-zero incoming wave at one point of the trap boundary $\delta\Omega$. Then the total wave
entering the trapping region the sum of the wave created by the
trapping condition and the incoming wave value
$\frac{1}{2}f(\delta\Omega^1,\cdot)+w^\pm(\delta\Omega^1,\cdot)$, where
$\delta\Omega^1$ denotes the first trap boundary reached. When
reaching the second trapping boundary $\delta\Omega^2$ the now
outgoing wave will see a matching force
$f(\delta\Omega^2,\cdot)=-\frac{1}{2}f(\delta\Omega^1,\cdot)$ leaving
an outgoing wave contribution
$w^\pm(\delta\Omega^2,\cdot)=w^\pm(\delta\Omega^1,\cdot)$ to escape the
trapping region $\Omega$.

In order to achieve selective radiation, only part of the content of a
trapped region are trapped at the boundary as can be achieved by using
a reduced force at the trapping boundary or by selectively omitting
certain phases in the trapping force pattern.

\subsection{Relationship of Traps to Non-Radiating Sources}

Marengo and Ziolkowski \cite{MZ99} present ideas very much related to
ideas presented here and in Berry {\em et al.}\cite{BFGW98}.

However, they arrive at a definition of non-radiating (NR) sources
that is not obviously similar to the traps presented here. In
particular they define NR sources as being non-interacting.  While
\cite{MZ99} note that a central property of NR sources is that they
store non-trivial field energy, traps described here can not only
store, but accumulate and selectively radiate waves.

The difference can be understood by observing that for example Berry
{\em et al.} assume a simple time-harmonic driver \cite[eq. (3)]{BFGW98}
throughout their discussion:

\begin{align}
f(x,t)=\text{Re}\left\{f(x)e^{-i\omega t}\right\}
\end{align}

By our earlier discussion we see that the temporal progression of the
boundary has to match the content of the interior domain. Hence once
the boundary is defined to be oscillatory the interior of the domain
needs to be spatially harmonic as derived in \cite{BFGW98,MZ99} and
has been rederived here. Hence a NR source as noted in literature,
with exception of the general orthogonality formulation for
time-varying sources given by Marengo, Devaney and Ziolkowski
\cite{MDZ99}, can be thought of as a time-oscillatory trap.

The arguments made here use a formalism that is discrete in
nature. However, the discreteness of the arguments are not necessarily
restrictive. The continuous case can be imagined with the discrete
time-step made small ($T\rightarrow 0$) or alternatively, discrete
pulses can be substituted with narrow distributions of
compact support. In neither case are the results of interest derived
here altered.

As has already been derived in \cite{BFGW98,GFW99} the critical
condition for non-propagating waves lie at the boundary of the
domain-range that the wave ought not to leave. In the discrete case it
is easy to see how this insight can be used and generalized. In fact,
the boundaries of the confining domain need not be static, nor need
the condition be used in a two-sided fashion.

\section{Conclusion}

In summary, this paper presented constructions of a broad class of
non-propagating sources on a string lattice model using trapping
conditions. In particular this includes numerical demonstrations of
finite-duration non-propagating excitations, directional excitations,
as well as virtually slowed waves. These examples help explain the
extension of non-propagating sources beyond the time-periodic case and
include treatment of onset, annihilation and spatial steering. These
properties ought to be observable in experiments well-described by the
wave equation. This equation often arises in problems in acoustics,
elasticity, optics and electromagneticsm. And hence the results presented here apply to these domains of application. While here I discussed the forward
problem, these results also relate to the inverse problem of finding
source contributions from the one-dimensional field state as occur for
example in acoustical, optical and electromagnetic detection problems.

\begin{acknowledgments}
I am grateful for reprints provided by Bruce Denardo and Greg Gbur,
who also brought relevant references to my attention. I'm also thankful for helpful comments and pointers to relevant literature by an anonymous referee.
\end{acknowledgments}

\appendix


\begin{thebibliography}{19}
\expandafter\ifx\csname natexlab\endcsname\relax\def\natexlab#1{#1}\fi
\expandafter\ifx\csname bibnamefont\endcsname\relax
  \def\bibnamefont#1{#1}\fi
\expandafter\ifx\csname bibfnamefont\endcsname\relax
  \def\bibfnamefont#1{#1}\fi
\expandafter\ifx\csname citenamefont\endcsname\relax
  \def\citenamefont#1{#1}\fi
\expandafter\ifx\csname url\endcsname\relax
  \def\url#1{\texttt{#1}}\fi
\expandafter\ifx\csname urlprefix\endcsname\relax\def\urlprefix{URL }\fi
\providecommand{\bibinfo}[2]{#2}
\providecommand{\eprint}[2][]{\url{#2}}

\bibitem[{\citenamefont{Berry et~al.}(1998)\citenamefont{Berry, Foley, Gbur,
  and Wolf}}]{BFGW98}
\bibinfo{author}{\bibfnamefont{M.}~\bibnamefont{Berry}},
  \bibinfo{author}{\bibfnamefont{J.~T.} \bibnamefont{Foley}},
  \bibinfo{author}{\bibfnamefont{G.}~\bibnamefont{Gbur}}, \bibnamefont{and}
  \bibinfo{author}{\bibfnamefont{E.}~\bibnamefont{Wolf}}, \bibinfo{journal}{Am.
  J. Phys.} \textbf{\bibinfo{volume}{66}}, \bibinfo{pages}{121}
  (\bibinfo{year}{1998}).

\bibitem[{\citenamefont{Denardo}(1998)}]{Denardo98}
\bibinfo{author}{\bibfnamefont{B.}~\bibnamefont{Denardo}},
  \bibinfo{journal}{Am. J. Phys.} \textbf{\bibinfo{volume}{66}},
  \bibinfo{pages}{1020} (\bibinfo{year}{1998}).

\bibitem[{\citenamefont{Gbur et~al.}(1999)\citenamefont{Gbur, Foley, and
  Wolf}}]{GFW99}
\bibinfo{author}{\bibfnamefont{G.}~\bibnamefont{Gbur}},
  \bibinfo{author}{\bibfnamefont{J.~T.} \bibnamefont{Foley}}, \bibnamefont{and}
  \bibinfo{author}{\bibfnamefont{E.}~\bibnamefont{Wolf}},
  \bibinfo{journal}{Wave Motion} \textbf{\bibinfo{volume}{30}},
  \bibinfo{pages}{125} (\bibinfo{year}{1999}).

\bibitem[{\citenamefont{Marengo and Ziolkowski}(1999)}]{MZ99}
\bibinfo{author}{\bibfnamefont{E.~A.} \bibnamefont{Marengo}} \bibnamefont{and}
  \bibinfo{author}{\bibfnamefont{R.~W.} \bibnamefont{Ziolkowski}},
  \bibinfo{journal}{Phys. Rev. Lett.} \textbf{\bibinfo{volume}{83}},
  \bibinfo{pages}{3345} (\bibinfo{year}{1999}).

\bibitem[{\citenamefont{Marengo and Ziolkowski}(2000{\natexlab{a}})}]{MZ00a}
\bibinfo{author}{\bibfnamefont{E.~A.} \bibnamefont{Marengo}} \bibnamefont{and}
  \bibinfo{author}{\bibfnamefont{R.~W.} \bibnamefont{Ziolkowski}},
  \bibinfo{journal}{J. Math. Phys.} \textbf{\bibinfo{volume}{41}},
  \bibinfo{pages}{845} (\bibinfo{year}{2000}{\natexlab{a}}).

\bibitem[{\citenamefont{Marengo and Ziolkowski}(2000{\natexlab{b}})}]{MZ00c}
\bibinfo{author}{\bibfnamefont{E.~A.} \bibnamefont{Marengo}} \bibnamefont{and}
  \bibinfo{author}{\bibfnamefont{R.~W.} \bibnamefont{Ziolkowski}},
  \bibinfo{journal}{Phys. Rev. E} \textbf{\bibinfo{volume}{62}},
  \bibinfo{pages}{4465} (\bibinfo{year}{2000}{\natexlab{b}}).

\bibitem[{\citenamefont{Marengo et~al.}(1999)\citenamefont{Marengo, Devaney,
  and Ziolkowski}}]{MDZ99}
\bibinfo{author}{\bibfnamefont{E.~A.} \bibnamefont{Marengo}},
  \bibinfo{author}{\bibfnamefont{A.~J.} \bibnamefont{Devaney}},
  \bibnamefont{and} \bibinfo{author}{\bibfnamefont{R.~W.}
  \bibnamefont{Ziolkowski}}, \bibinfo{journal}{J. Opt. Soc. Am. A}
  \textbf{\bibinfo{volume}{16}}, \bibinfo{pages}{1612} (\bibinfo{year}{1999}).

\bibitem[{\citenamefont{Marengo and Ziolkowski}(2000{\natexlab{c}})}]{MZ00b}
\bibinfo{author}{\bibfnamefont{E.~A.} \bibnamefont{Marengo}} \bibnamefont{and}
  \bibinfo{author}{\bibfnamefont{R.~W.} \bibnamefont{Ziolkowski}},
  \bibinfo{journal}{IEEE Trans. Antennas Propag.}
  \textbf{\bibinfo{volume}{48}}, \bibinfo{pages}{1553}
  (\bibinfo{year}{2000}{\natexlab{c}}).

\bibitem[{\citenamefont{Marengo et~al.}(2000)\citenamefont{Marengo, Devaney,
  and Ziolkowski}}]{MDZ00}
\bibinfo{author}{\bibfnamefont{E.~A.} \bibnamefont{Marengo}},
  \bibinfo{author}{\bibfnamefont{A.~J.} \bibnamefont{Devaney}},
  \bibnamefont{and} \bibinfo{author}{\bibfnamefont{R.~W.}
  \bibnamefont{Ziolkowski}}, \bibinfo{journal}{J. Opt. Soc. Am. A}
  \textbf{\bibinfo{volume}{17}}, \bibinfo{pages}{34} (\bibinfo{year}{2000}).

\bibitem[{\citenamefont{Hoenders and Ferwerda}(2001)}]{HF01}
\bibinfo{author}{\bibfnamefont{B.~J.} \bibnamefont{Hoenders}} \bibnamefont{and}
  \bibinfo{author}{\bibfnamefont{H.~A.} \bibnamefont{Ferwerda}},
  \bibinfo{journal}{Phys. Rev. Lett.} \textbf{\bibinfo{volume}{87}},
  \bibinfo{pages}{060401} (\bibinfo{year}{2001}).

\bibitem[{\citenamefont{Denardo and Miller}(2003)}]{DM03}
\bibinfo{author}{\bibfnamefont{B.}~\bibnamefont{Denardo}} \bibnamefont{and}
  \bibinfo{author}{\bibfnamefont{G.~L.} \bibnamefont{Miller}},
  \bibinfo{journal}{Am. J. Phys.} \textbf{\bibinfo{volume}{71}},
  \bibinfo{pages}{778} (\bibinfo{year}{2003}).

\bibitem[{\citenamefont{Gbur}(2003)}]{Gbur03}
\bibinfo{author}{\bibfnamefont{G.}~\bibnamefont{Gbur}}, in
  \emph{\bibinfo{booktitle}{Progress in Optics}}, edited by
  \bibinfo{editor}{\bibfnamefont{E.}~\bibnamefont{Wolf}}
  (\bibinfo{publisher}{Elsevier}, \bibinfo{address}{Amsterdam},
  \bibinfo{year}{2003}), vol.~\bibinfo{volume}{45}, pp.
  \bibinfo{pages}{273--315}.

\bibitem[{\citenamefont{Schott}(1937)}]{Schott37}
\bibinfo{author}{\bibfnamefont{G.~A.} \bibnamefont{Schott}},
  \bibinfo{journal}{Proc. R. Soc. London, Ser. A}
  \textbf{\bibinfo{volume}{159}}, \bibinfo{pages}{570} (\bibinfo{year}{1937}).

\bibitem[{\citenamefont{Schott}(1933)}]{Schott33}
\bibinfo{author}{\bibfnamefont{G.~A.} \bibnamefont{Schott}},
  \bibinfo{journal}{Phil. Mag.} \textbf{\bibinfo{volume}{15}}
  (\bibinfo{year}{1933}).

\bibitem[{\citenamefont{Bohm and Weinstein}(1948)}]{BW48}
\bibinfo{author}{\bibfnamefont{D.}~\bibnamefont{Bohm}} \bibnamefont{and}
  \bibinfo{author}{\bibfnamefont{M.}~\bibnamefont{Weinstein}},
  \bibinfo{journal}{Phys. Rev.} \textbf{\bibinfo{volume}{74}},
  \bibinfo{pages}{1789} (\bibinfo{year}{1948}).

\bibitem[{\citenamefont{Goedecke}(1964)}]{Goedecke64}
\bibinfo{author}{\bibfnamefont{G.~H.} \bibnamefont{Goedecke}},
  \bibinfo{journal}{Phys. Rev.} \textbf{\bibinfo{volume}{135}},
  \bibinfo{pages}{B281} (\bibinfo{year}{1964}).

\bibitem[{\citenamefont{Habashy et~al.}(1990)\citenamefont{Habashy, Chow, and
  Dudley}}]{HCD90}
\bibinfo{author}{\bibfnamefont{T.~M.} \bibnamefont{Habashy}},
  \bibinfo{author}{\bibfnamefont{E.~Y.} \bibnamefont{Chow}}, \bibnamefont{and}
  \bibinfo{author}{\bibfnamefont{D.~G.} \bibnamefont{Dudley}},
  \bibinfo{journal}{IEEE Trans. Antennas Propag.}
  \textbf{\bibinfo{volume}{38}}, \bibinfo{pages}{668} (\bibinfo{year}{1990}).

\bibitem[{\citenamefont{Smith}(1997)}]{Smith97}
\bibinfo{author}{\bibfnamefont{J.~O.} \bibnamefont{Smith}}, in
  \emph{\bibinfo{booktitle}{Musical Signal Processing}}, edited by
  \bibinfo{editor}{\bibfnamefont{C.}~\bibnamefont{Roads}},
  \bibinfo{editor}{\bibfnamefont{S.~T.} \bibnamefont{Pope}},
  \bibinfo{editor}{\bibfnamefont{A.}~\bibnamefont{Piccialli}},
  \bibnamefont{and} \bibinfo{editor}{\bibfnamefont{G.}~\bibnamefont{De~Poli}}
  (\bibinfo{publisher}{Swets}, \bibinfo{address}{Lisse, Netherlands},
  \bibinfo{year}{1997}), chap.~\bibinfo{chapter}{7}, pp.
  \bibinfo{pages}{221--263}.

\bibitem[{\citenamefont{Kreyszig}(1999)}]{Kreyszig99}
\bibinfo{author}{\bibfnamefont{E.}~\bibnamefont{Kreyszig}},
  \emph{\bibinfo{title}{{Advanced Engineering Mathematics}}}
  (\bibinfo{publisher}{John Wiley \& Sons}, \bibinfo{address}{New York},
  \bibinfo{year}{1999}), \bibinfo{edition}{8th} ed.

\end{thebibliography}

\begin{figure}[bp]
\centering
\caption{Simulation of a non-propagating excitation of width $3$ which
is annihilated after $3.5$ periods. The total temporal length of the excitation is $10$. The excitation leaves the string at rest after it is completed. Top: Complete wave pattern. Bottom: Excitation only.}
\label{fig:nonprop}
\end{figure}

\begin{figure}[bp]
\centering
\caption{Simulation of a non-propagating excitation of width $3$ showing escaping waves at the onset transient. Top: Complete wave pattern. Bottom: Excitation only.}
\label{fig:pointpair}
\end{figure}

\begin{figure}[bp]
\centering
\caption{Simulation of a directional excitation of width $3$. The
deflected component experienced a sign inversion. The temporal length
of the excitation sequence is two, including the initial impulse. Top: Complete wave. Bottom: Excitations only.}
\label{fig:onesided}
\end{figure}

\begin{figure}[bp]
\centering
\caption{Simulation of a finite-duration virtual slow wave excitation of width $3$. The wave is annihilated after 10 steps. Top: Complete wave. Bottom: Excitation only.}
\label{fig:slowwave}
\end{figure}

\begin{figure}[bp]
\centering
\caption{A grazing propagating wave against a changing trap boundary can create regions (gray) in which no trap affect applies.}
\label{fig:steering}
\end{figure}

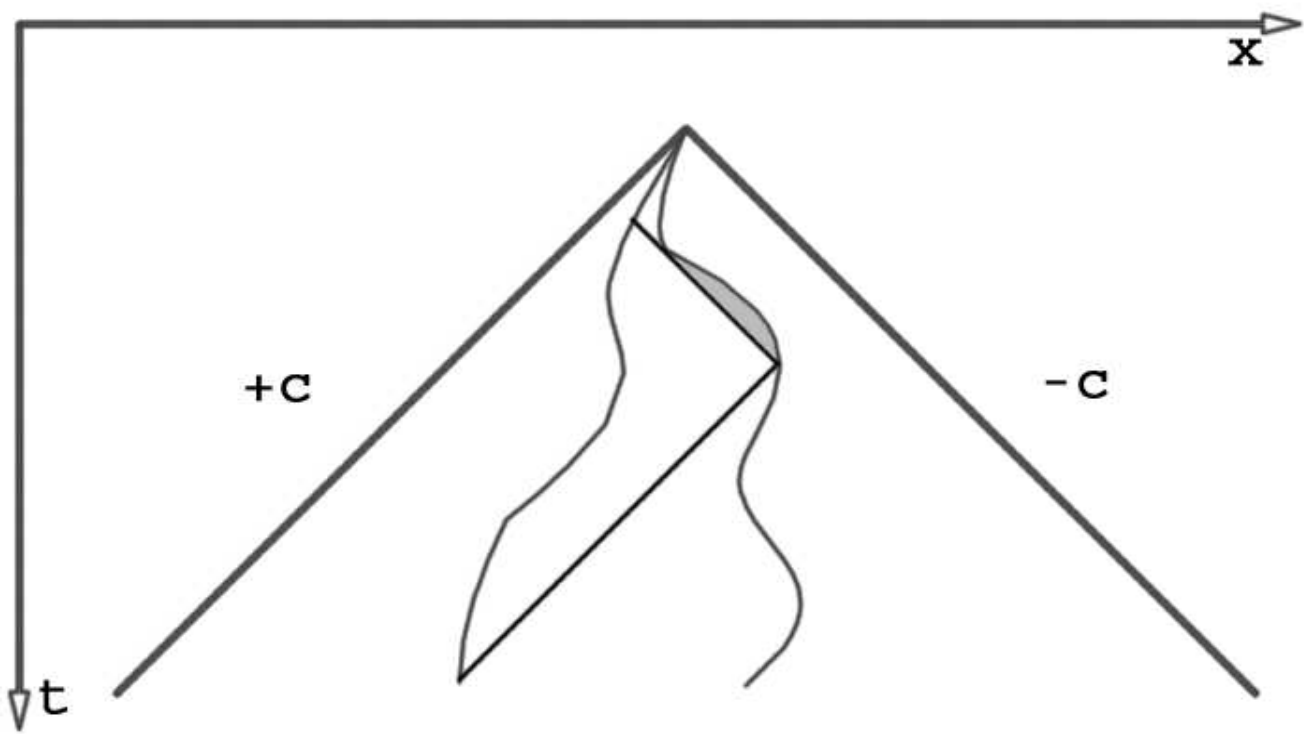
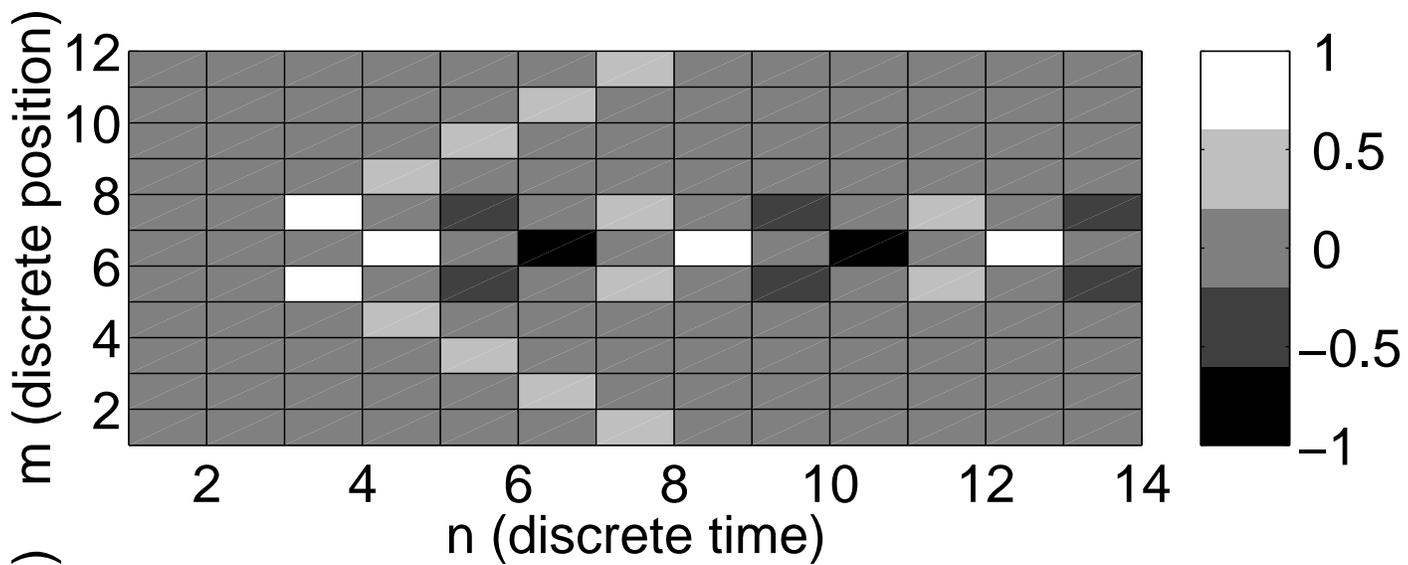
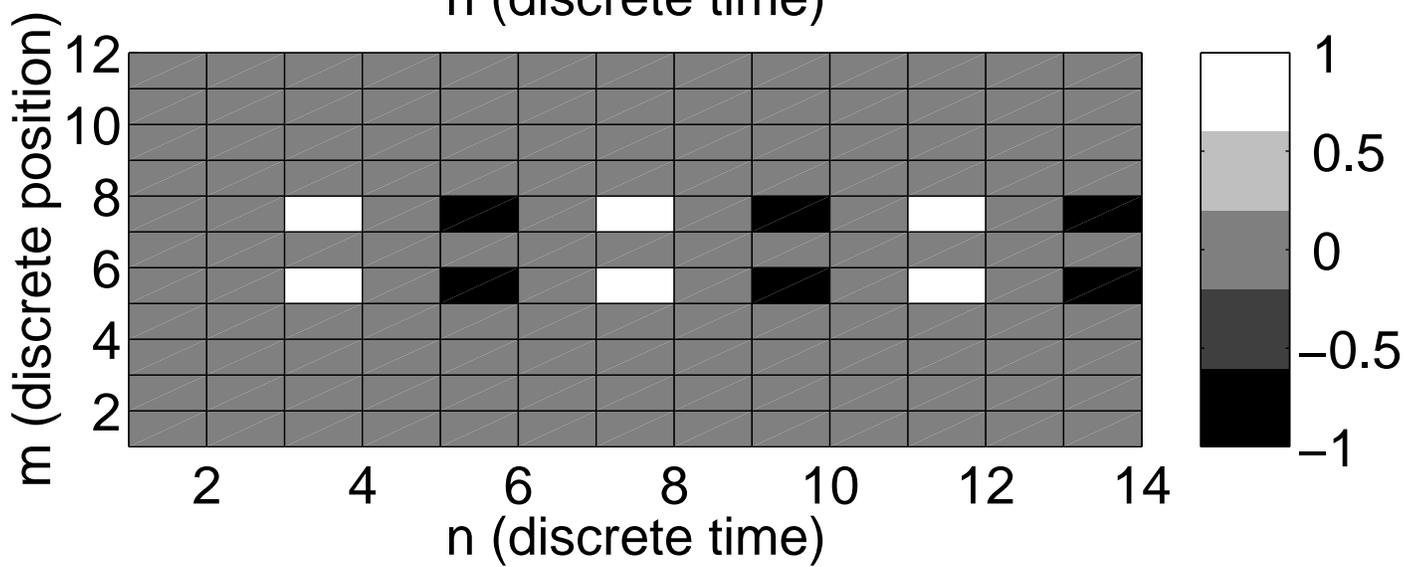
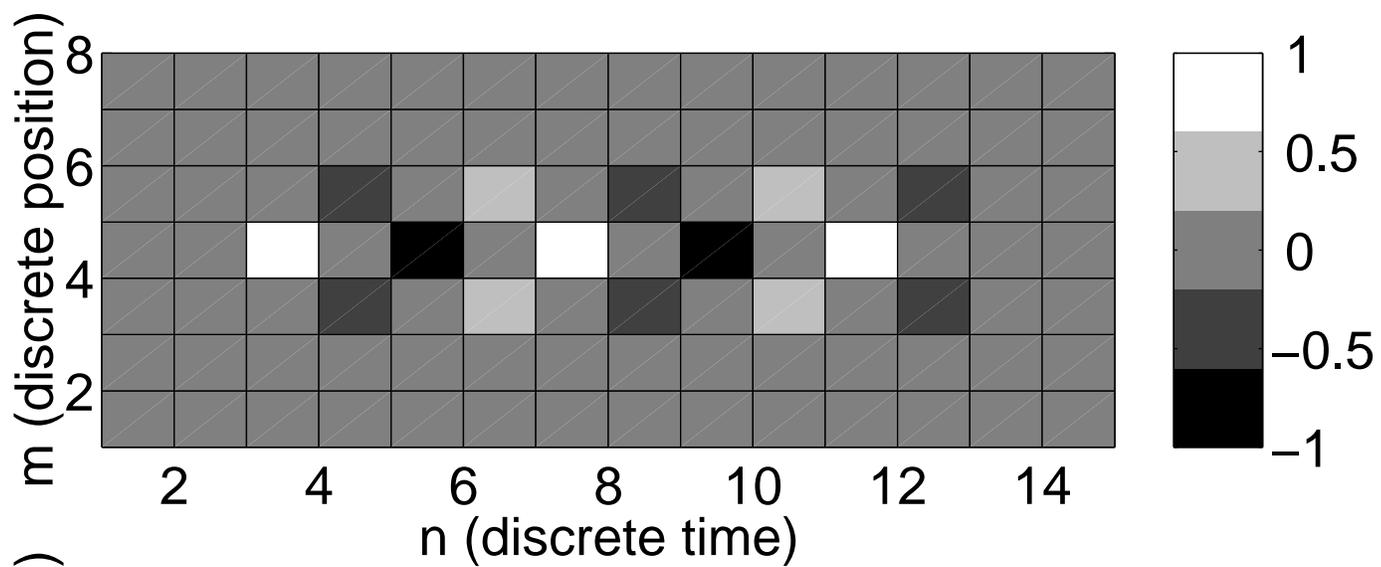
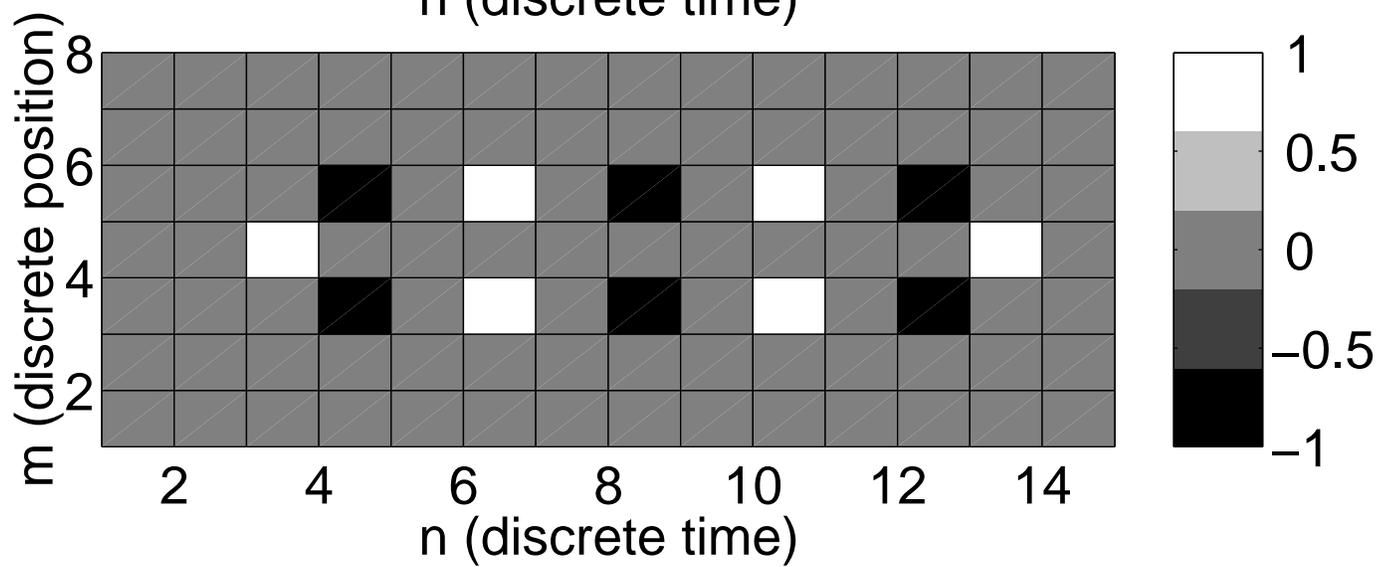
\end{document}